# Rapid and Accurate Detection of SARS-CoV-2 Mutations using a Cas12a-based Sensing Platform

Changsheng He[1], Cailing Lin[1], Guosheng Mo[1], Binbin Xi[1], An'an Li[3], Dongchao Huang[1], Yanbin Wan[1], Feng Chen[1], Yufeng Liang[3], Qingxia Zuo[1], Wanqing Xu[1], Dongyan Feng[1], Guanting Zhang[4], Liya Han[1], Changwen Ke[2], Hongli Du[1,*], Lizhen Huang[1,*]

[1] School of Biology and Biological Engineering, South China University of Technology, Guangzhou, 510006, China.

[2] Guangdong Provincial Center for Disease Control and Prevention, Guangzhou 511430, China.

[3] School of public health, Southern Medical University, Guangzhou, 510515, China.

[4] School of public health, Sun Yat-sen University, Guangzhou, 510080, China.

* Corresponding Emails: huanglzh@scut.edu.cn; hldu@scut.edu.cn

**Abstract**

The increasing prevalence of SARS-CoV-2 variants with spike mutations has raised concerns owing to higher transmission rates, disease severity, and escape from neutralizing antibodies. Rapid and accurate detection of SARS-CoV-2 variants provides crucial information concerning the outbreaks of SARS-CoV-2 variants and possible lines of transmission. This information is vital for infection prevention and control. We used a Cas12a-based RT-PCR combined with CRISPR on-site rapid detection system (RT-CORDS) platform to detect the key mutations in SARS-COV-2 variants, such as 69/70 deletion, N501Y, and D614G. We used type-specific CRISPR RNAs (crRNAs) to identify wild-type (crRNA-W) and mutant (crRNA-M) sequences of SARS-CoV-2. We successfully differentiated mutant variants from wild-type SARS-CoV-2 with a sensitivity of $10^{-17}$ M (approximately 6 copies/μL). The assay took just 10 min with the Cas12a/crRNA reaction after a simple RT-PCR using a fluorescence reporting system. In addition, a sensitivity of $10^{-16}$ M could be achieved when lateral flow strips were used as readouts. The accuracy of RT-CORDS for SARS-CoV-2 variant detection was 100% consistent with the sequencing data. In conclusion, using the RT-CORDS platform, we accurately, sensitively, specifically, and rapidly detected SARS-CoV-2 variants. This method may be used in clinical diagnosis.

## 1. Introduction

A number of severe acute respiratory syndrome coronavirus 2 (SARS-CoV-2) variants have emerged since its first outbreak in 2019 (Sanyaolu et al. 2021). Some of these variants (termed as Alpha/Beta/Gamma/Delta), which have been reported to influence viral transmissibility, disease severity, diagnostics, and vaccine efficacy, were defined as the variants of concern (VOC) by the World Health Organization (WHO) (World Health Organization. 2021). The emergence of VOC has reported in > 190 countries and regions, and > 68% of global samples are positive for VOC (Table S1).

Because the spike protein mediates SARS-COV-2 entry into the host and is also a significant target of neutralizing antibodies and vaccines, some key mutations of VOC occurring in the spike protein (e.g., D614G, N501Y and 69/70 deletion) have garnered significant scientific attention (Harvey et al. 2021; Salvatori et al. 2020). D614G, one of the key mutations, is a nonsynonymous amino acid change in the spike protein due to a 23,403 A>G single-base substitution in the SARS-CoV-2 reference sequence; the mutation was first identified in early 2020 (Korber et al. 2020). It is found in four SARS-CoV-2 VOC: B.1.1.7, P1, P1.351, and B.1.617.2. Patients with G614 COVID-19 show higher viral loads and 61% higher mortality than those with COVID-19 due to pre-existing variants (Davies et al. 2021a; Davies et al. 2021b; Korber et al. 2020; Plante et al. 2020). In addition, D614G variants have been demonstrated to increase the transmissibility in *in vitro* human cell models (from 1.3 to 7.7 folds) and *in vivo* animal models. The D614G mutation may alter proteolytic cleavage and further increase S1 shedding, or it could enhance the binding affinity between RBD and AEC2 (Daniloski et al. 2021; Hou et al. 2020; Plante et al. 2020; Wang et al. 2021a; Zhang et al. 2020).

N501Y is also a nonsynonymous amino acid change due to 23,063 A>T substitution in the SARS-CoV-2 reference sequence. This mutation is identified in various VOC such as P1, B.1.1.7, and B.1.351 lineages. Clinical data indicate that N501Y-positive VOC significantly increase the risks of hospitalization (59% higher), intensive care unit admission (105% higher), and death (61% higher) (Fisman and Tuite 2021). Recent evidence indicates that the N501Y mutation enhances the affinity to bind with the host angiotensin-converting enzyme 2 (ACE2) receptor by up to 7-fold more than that of the wild-type (WT) (Ali et al. 2021; Laffeber et al. 2021; Li et al. 2021b; Liu et al. 2021) and also enhances viral transmission both *in vivo* (1.0–5.3 fold) and *in vitro* (1.3–5.4 fold) (Liu et al. 2021). Furthermore, the N501Y mutation exhibits poorer binding potential for 82.7% of the common alleles of the major histocompatibility complex than N501 and might escape immune defenses (Castro et al. 2021). The 69/70 deletion results from 204–209del in the S gene and has been found in both B.1.1.7 and B.1.526.1 variants. This small deletion is associated with the S-gene target failure of a three-target reverse transcription polymerase chain reaction (RT-PCR) assay (Bal et al. 2021). Recently, a 2-fold increase in pseudovirus infectivity of 69/70 deletion was observed in several cell models of ACE2 expression (Meng et al. 2021a). Therefore, owing to the prominent impacts on viral transmissibility and disease severity, SARS-CoV-2 VOC or other variants with key mutations should be monitored and identified for enhanced epidemic control and improved clinical therapy.

Until now, sequencing has been the gold standard for identifying SARS-CoV-2 variant mutations. This method is accurate but time-consuming and costly and must be performed off-site. Moreover, quantitative RT-PCR (RT-qPCR) has been developed as a gold standard for detection of SARS-CoV-2 infection, identification of SARS-CoV-2 mutations with mutation-matched primers or probes, and identification through amplification and melting curve analyses (Aoki et al. 2021; Bedotto et al. 2021; Vega-Magaña et al. 2021; Zelyas et al. 2021). However, primer/probe mismatches in a qPCR may only slightly alter these curves (Boyle et al. 2009; Stadhouders et al. 2010; Süß et al. 2009). Thus, the PCR-based identification of single-base mutations is nonspecific and not sufficiently reliable. In addition, performing quantitative RT-PCR assays requires professional operations and bulky instruments.

Cas12/crRNA method, which is different from the traditional nucleic acid detection methods, recognizes targets highly specifically with only 18–20-nt crRNA; this method has been widely used for the rapid diagnosis of various viruses such as African swine fever virus (AFSV), hepatitis B virus (HBV), and SARS-CoV-2 (Bai et al. 2019; Ding et al. 2021; Ma et al. 2021; Wang et al. 2021b). By introducing an additive mismatch in crRNA, the CRISPR/Cas12 system can even be used for single-nucleotide polymorphism (SNP) genotyping with single-base specificity (Chen et al. 2021; Huang et al. 2021; Lee et al. 2020; Lee Yu et al. 2021; Meng et al. 2021b). The identification of a single base makes Cas12 an ideal approach for mutation detection.

Therefore, we assessed the detection of SARS-COV-2 mutations of concern (69/70 deletion, N501Y, and D614G) using an improved Cas12a-based CRISPR on-site rapid detection system (CORDS) platform—a rapid, sensitive, and specific on-site biosensing system developed in our laboratory. We designed and screened mutant (MT)-specific crRNA (crRNA-M). Cas12/crRNA-M could specifically and rapidly differentiate SARS-CoV-2 mutations compared with WT-specific crRNA (crRNA-W). The sensitivity was $10^{-17}$ M (6 copies/µL) with a fluorescence reporting system and up to $10^{-16}$ M (60 copies/µL) using a lateral flow strip reporting system. The accuracy of RT-CORDS is 100% consistent with that of the sequencing method. These findings indicate that the improved RT-CORDS platform is a powerful tool for monitoring the key mutations in VOC and other SARS-CoV-2 variants.

## 2. Materials and Methods

### 2.1 Materials

LbCas12a, HiScribe T7 High-Yield RNA Synthesis Kit, RNA Cleanup Kit, NEBNext Q5 Hot Start HiFi PCR Master Mix, NEBuffer2.1 were ordered from NEB (Beijing, China). DNaseI, recombination RNase inhibitor (RRI), PrimeSTAR max were ordered from Takara Bio (Beijing, China). T-vectors, Pfu DNA polymerase were ordered from TIANGEN Biotechnology (Beijing, China). One-Step RT-PCR Kit was ordered from Vazyme Biotech (Nanjing, China). Gel Extraction Kit was ordered from Omega Biotech (Shanghai, China). Lateral flow strips were purchased from Magigen Biotech (Guangzhou, China). Oligonucleotides were synthesized by GENEWIZ (Jiangsu, China). *S* gene DNA targets were synthesized by Generay Biotech (Shanghai, China). Nucleic acid was quantified using Thermo Fisher Nanodrop 1000 Spectrophotom. Fluorescence signals were recorded with Tecan's Spark 20M.

## 2.2 Design and transcription of crRNAs

We designed the crRNAs for mutation identification based on a previous study (Bai et al. 2019). In brief, an 18-nt sequence following the protospacer adjacent motif (PAM; TTTV), which covers the mutation, was selected as the crRNA. In addition, as a single-base mutation of D614G and N501Y, an additional mismatch was introduced in the seed region of their crRNA-W and crRNA-M to enhance the differentiability between the WT and MT variants (Table S2). For the 69/70 target, TTC was selected as the PAM and no additional mismatch was introduced to the crRNA-W/M because this mutation was small deletion instead of single-base substitution (Table S2).

crRNA transcription templates were prepared using 60-nt oligos containing a T7 promoter, scaffold (Table S3). After oligo-F and oligo-R annealing, the annealed products were ligated to T-vectors to obtain pGM-T-crRNA plasmids. Finally, transcription templates were obtained through PCR amplification with Pfu DNA polymerase from pGM-T-crRNA and were purified using a Gel Extraction Kit.

*In vitro* transcription of crRNAs used a HiScribe T7 High-Yield RNA Synthesis Kit. Reactions were performed in 20 μL volume at 37°C for 16 h following the manufacturer's instruction for short RNA transcripts. Finally, the crRNA transcripts were treated and purified with an RNA Cleanup Kit.

## 2.3 Preparation of WT and MT SARS-CoV-2 DNA and RNA targets

The DNA targets of WT S gene and MT SARS-CoV-2 (covering D614G, N501Y and 69/70 deletion sites) were synthesized (Table S4) and then cloned into pUC57 plasmid.

RNA targets were prepared through in vitro transcription with a HiScribe T7 High-Yield RNA Synthesis Kit. Briefly, RNA transcription templates containing a T7 promoter were amplified from pUC57-WT-DNA and pUC57-MT-DNA with specific primers (Table S2) using the NEBNext Q5 Hot Start HiFi PCR Master Mix according to manufacturer's instruction, followed by purification using a Gel Extraction Kit. *In vitro* transcription was performed in 20 μL reaction volume, according to a standard RNA synthesis protocol, at 37°C for 16 h. Finally, RNA transcripts were treated with DNaseI to remove transcription templates, purified with an RNA

Cleanup Kit, and quantified using Thermo Fisher Nanodrop 1000 Spectrophotom.

**2.4 *In vitro* cleavage assays**

Briefly, LbCas12a-mediated target cleavage assays were performed in 20 μL reaction volume with 50 nM LbCas12a protein, 100 nM crRNA, 1× NEBuffer2.1, 20 U recombination RNase inhibitor (RRI), 5 nM linear pUC57-WT-DNA or pUC57-MT-DNA plasmids, and nuclease-free water. LbCas12a was preincubated with crRNA and RRI in NEBuffer2.1 at room temperature for 10 min to form ribonucleoproteins. Target DNA was then added, and the reaction mixture was incubated at 37°C for 1 or 2 h. Finally, cleaved products were verified using gel electrophoresis.

**2.5 Modified CORDS assay**

Bioinformatic alignment of multiple SARS-CoV-2 sequences downloaded from GISAID was performed to identify specific and relatively conserved sequences adjacent to mutations. These sequences were used to design PCR/RT-PCR primers (TableS2).

The modified CORDS fluorescence assay comprises two steps, PCR amplification and Cas12a sensing. PCR reactions were performed in 50 μL reaction volume containing 20 μL of DNA, 4 μL of 10 μM primer F/R mix, 25 μL of 2× PrimeSTAR max, and nuclease-free water. Thermal cycling was as follows: 3 min at 98°C for initial denaturation; 30 cycles of 10 s at 98°C for denaturation, 10 s at 55°C for annealing, and 5 s at 72°C for extension; and 5 min at 72°C for the final extension. For Cas12a reaction, FAM- and BHQ-labeled 12-nt ssDNA reporter (FQ-ssDNA) (FAM-NNNNNNNNNNNN-BHQ) for reporting the collateral cleavage of Cas12a was synthesized. The fluorescence reporting assay was performed following a detection protocol. The whole 20 μL reaction volume contained 50 nM LbCas12a, 100 nM crRNA, and 20 U RNase inhibitor, 1× NEBuffer2.1, 10 μL PCR product, 500 nM FQ-ssDNA, and nuclease-free water at room temperature. The reaction mixture was then quickly transferred to a 384-well fluorescence plate reader and incubated at 37°C. Fluorescence signals were detected using Tecan's Spark 20M multimode microplate reader at an excitation wavelength of 485 nm and an emission wavelength of 535 nm. Fluorescence assay kinetics assay were assessed with signals collected every 2.5 or 5 min.

The PCR reaction for the CORDS lateral flow strip assay was performed as described earlier in this section. Digoxin- and biotin-labeled 14-nt ssDNA reporter (DB-ssDNA) (Digoxin-NNNNNNNNNNNNNN-Biotin) was used to report Cas12a collateral activity in the lateral flow strip reporting assay. Briefly, a Cas12a paper strip assay was performed in 40 μL reaction volume containing nuclease-free water, 1× NEBuffer2.1, 40 U RNase inhibitor, 50 nM LbCas12a, 100 nM crRNA, 2 nM DB-ssDNA, and 20 μL PCR product. The reaction mixture was then incubated at 37°C for 60 min. Finally, lateral flow strips were inserted into the reaction mixture and incubated at room temperature for 5 min for use as readouts.

## 2.6 RT-CORDS assay

The RT-CORDS fluorescence assay comprises an RT-PCR reaction for amplification and a Cas12a reaction for sensing. A One-Step RT-PCR Kit was used. In brief, 50μL reaction volume contains 18.5 μL of RNA sample, 4 μL of 10 μM primer F/R mix, 25 μL of 2× One Step mix and 2.5 μL of One Step Enzyme mix. The amplification program was as follows: 30 min at 50°C for reverse transcription; 3 min at 94°C for initial denaturation; 30 cycles of 30 s at 94°C for denaturation, 30 s at 55°C for annealing, and 30 s at 72°C for extension; and 5 min at 72 °C for the final extension. The Cas12a reaction was performed using a method similar to that used for the CORDS fluorescence assay, except the RT-PCR product substituted the PCR product.

The RT-PCR reaction for the RT-CORDS lateral strip assay was performed as described earlier in this section. The Cas12a reaction was performed using a method similar to that used for the CORDS lateral flow strip assay, except the RT-PCR product substituted the PCR product.

## 2.7 SARS-CoV-2 variant detection

RNA samples of 18 SARS-CoV-2 variants were collected from Guangdong Provincial Center for Disease Control and Prevention (CDC) and treated in strict accordance with the WHO-recommended procedure and were used in the RT-CORDS assay to detect real SARS-CoV-2 variants. All samples with N501Y, D614G, and 69/70 deletion mutations were detected. RT-PCR products of the samples were also sequenced for further validation.

**2.8 Lyophilization of RT-CORDS**

We lyophilized the Cas12a lateral flow strip reporting system to simplify the detection system for operation and production. The lyophilized system comprises two tubes: W and M. The W-tube contains NEBuffer2.1, RNase inhibitor, Cas12a, crRNA-W, and ssDNA DB reporters. This mixture was prefrozen at −80°C for 12 h and freeze-dried at −50°C for 4 h. The M-tube contains crRNA-M freeze-dried following the abovementioned process.

The lyophilized assay was performed by simply redissolving W- and M-tube contents in 20 μL nuclease-free water and then adding 20 μL RT-PCR products at room temperature. Reactions are then performed as described above. Finally, lateral flow strips are inserted into the reaction mixture and incubated at room temperature for use as readouts.

**2.9 Statistical Analysis**

All the replicate experiments in this study consisted of three repeats. Uncertainties in mean values are provided as standard errors. Statistical analyses were performed using GraphPad Prism. Statistical significance was assessed using 2-way analysis of variance and multiple comparisons. Significance was considered at *$p < 0.05$, **$p < 0.01$, ***$p < 0.001$, ****$p < 0.0001$; ns indicates no significance.

**3. Results**

**3.1 Highly specific SARS-CoV-2 mutation and cleavage by modified crRNA**

The SARS-CoV-2 reference genome sequence was downloaded from the National Center for Biotechnology Information and mutation information for SARS-CoV-2 variants was obtained from GISAID (https://www.gisaid.org/). Because CRISPR/Cas12a has been reported to differentiate SNPs with single-base resolution (Chen et al. 2021; Lee Yu et al. 2021; Li et al. 2018), we used Cas12a to identify SARS-CoV-2 N501Y, D614G, and 69/70 deletion mutations; these are the key mutations of SARS-CoV-2 VOC (Table S1). These mutations are associated with transmissibility and clinical characteristics of SARS-CoV-2 variants (Hou et al. 2020; Laffeber et al. 2021; Meng et al. 2021a). We synthesized crRNA-W and crRNA-M to target WT and MT sequences for each mutation and performed *in vitro* cleavage assays to verify the ability of the CRISPR/Cas12a system to identify SARS-CoV-2 mutations (Table S1, Figs. 1A – 1D).

We designed 501-crRNA-W (N501-WT-specific crRNA) and 501-crRNA-M (Y501-MT-specific crRNA) to detect N501Y mutation (Table S2). The kinetic profiling of single and double mismatches between the crRNA and target (Jones et al. 2021), suggested the introduction of a skillful mismatch to the seed region of both 501-crRNA-W and 501-crRNA-M that make MTs easier to differentiate from WTs. As expected, we found that N501-WT substrate was significantly cleaved in the presence of 501-crRNA-W, but no cleavage was observed in the presence of 501-crRNA-M. Likewise, the 501-MT substrate was cleaved significantly by 501-crRNA-M compared with 501-crRNA-W (Fig. 1B). The abovementioned findings indicate that 501-crRNA-W and 501-crRNA-M can identify N501Y MT variants specifically.

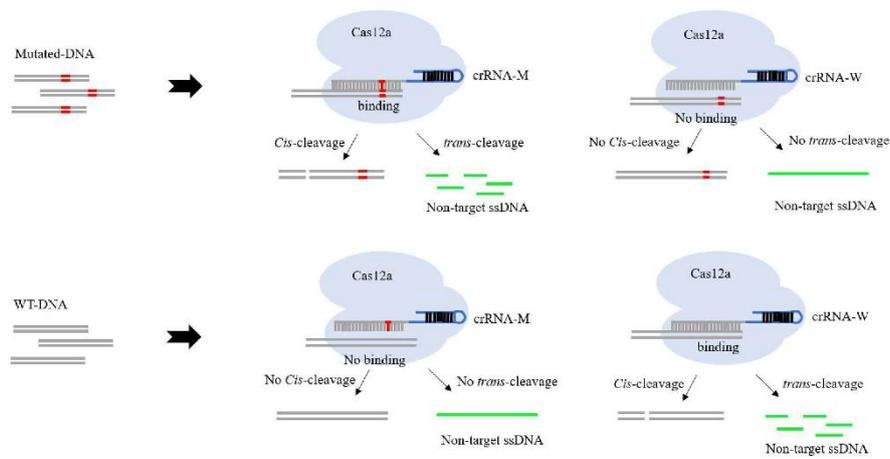

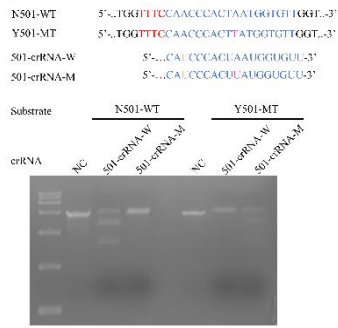 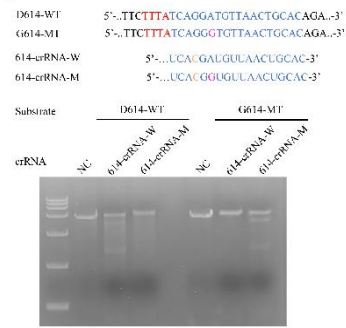 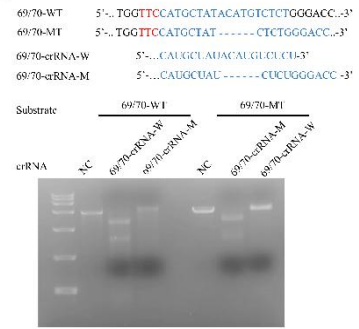

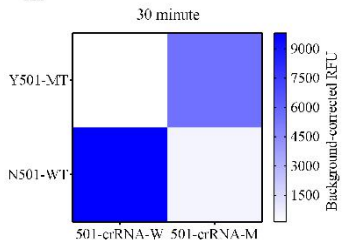 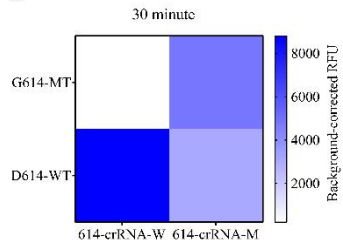 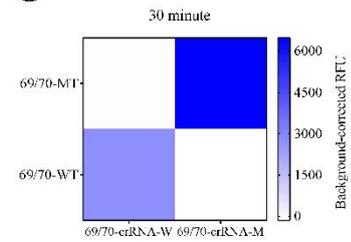

Fig.1. Specific identification of SARS-CoV-2 mutations using CRISPR/Cas12a *cis*-cleavage and *trans*-cleavage by MT-specific crRNA-M and WT-specific crRNA-W. (A) Schematic representation of Cas12a/crRNA-W and Cas12a/crRNA-M to differentiate MT-DNA from WT-DNA. (B, C, D) in vitro cleavage assay of N501Y, D614G and 69/70 deletion targets using Cas12a/crRNA-W and Cas12a-crRNA-M. WT: wild-type DNA; MT: mutant DNA; NC: negative control. (E, F, G) Identification of N501Y, D614G and 69/70 deletion targets through CRISPR/Cas12a *trans*-cleavage of the fluorescence reporter.

We also introduced a skillful mismatch into the seed region of 614-crRNA-W and 614-crRNA-M for D614G. The targets were respectively recognized and cleaved specifically and efficiently by 614-crRNA-W and 614-crRNA-M (Fig. 1C). However, no canonical PAM TTTV motif was present adjacent to the 69/70 mutation site. We selected TTC as the PAM for LbCas12a recognition (Table S2); LbCas12a can also cleave targets with this PAM (Yamano et al. 2017). We designed 69/70-crRNA-W and 69/70-crRNA-M to specifically target 69/70-WT and 69/70-MT without introducing any artificial mismatch. 69/70-WT substrate was only cleaved by 69/70-crRNA-W, and 69/70-MT substrate was cleaved by only 69/70-crRNA-M (Fig. 1D). We conclude that the specific crRNA-W and crRNA-M that we designed can successfully and specifically identify N501Y, D614G, and 69/70 deletion mutations in SARS-CoV-2 variants.

**3.2 CORDS can accurately and rapidly differentiate the DNA targets of SARS-CoV-2 mutations**

We have previously established a CORDS platform for *in vitro* virus nucleic acid detection (Bai et al. 2019). For mutation identification, we also introduced fluorescence probes, 12-nt ssDNAs labeled with 5′-FAM and 3′-BHQ, to report the collateral activity of Cas12a. We assessed the specificity of mutation discrimination by integrating a fluorescence reporter system. Fluorescence increased solely when 501-crRNA-M was used to detect Y501-MT substrate (Fig. 1E, S1A, and S1B). The same results were observed when the substrate was N501-WT, 69/70-WT, or 69/70-MT (Figs. 1F, 1G, and S1C–S1F). The G614-MT substrate could only activate Cas12a/614-crRNA-M but not Cas12a/614-crRNA-W. D614-MT can be slightly activated by Cas12a/614-crRNA-W (Figs. 1F and S1D). However, significant differences in fluorescence intensities between 614-crRNA-W and 614-crRNA-M were noted, and this slight activation had no impact on D614G discrimination. Furthermore, kinetic data showed that 30 min is enough time to differentiate the MT variants from the WT (Figs. 1E–1G).

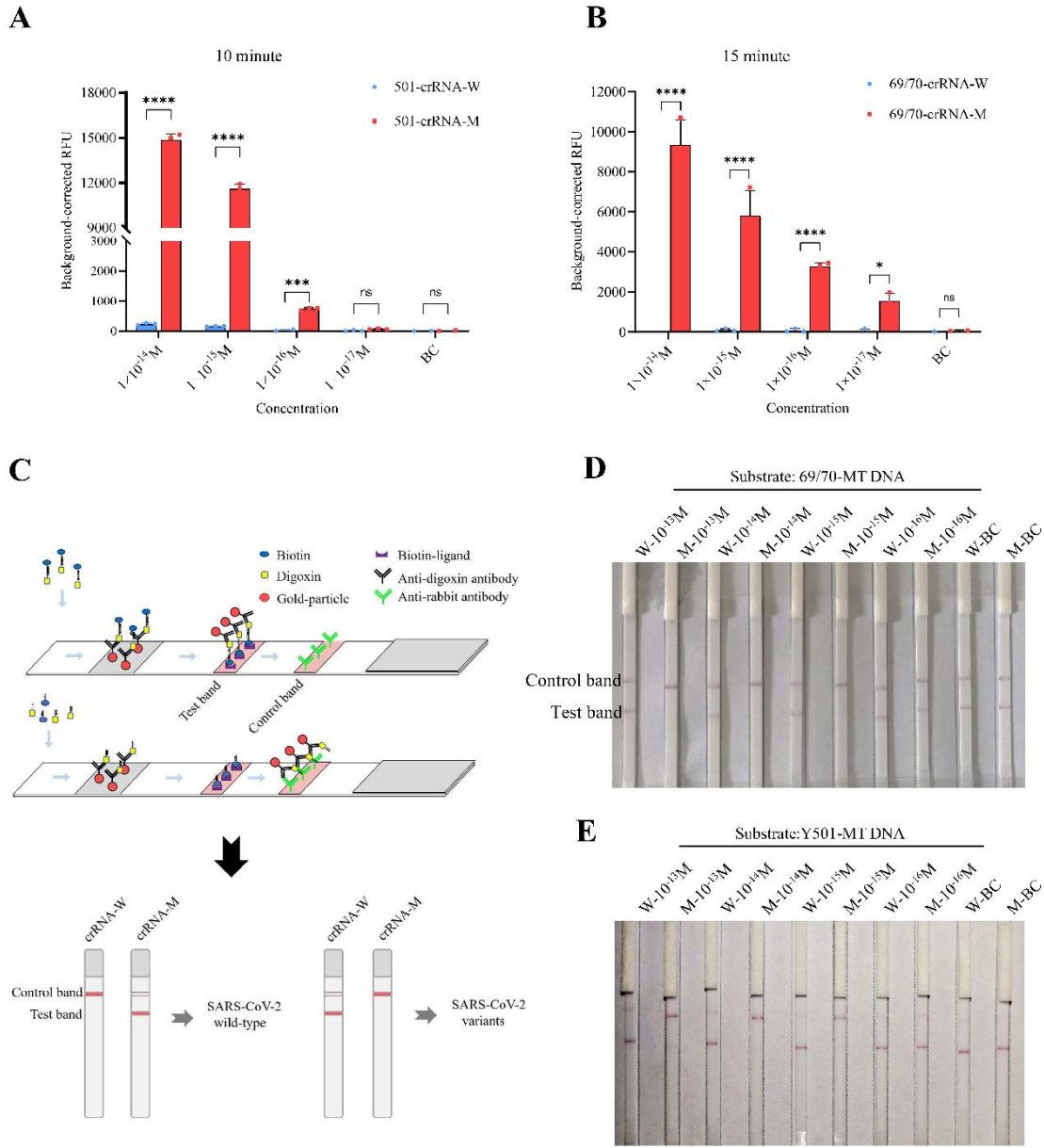

Fig. 2. Detection of N501Y and 69/70 deletion mutations in SARS-CoV-2 using the CORDS platform. (A) Sensitivity of N501Y detection using the CORDS fluorescence reporting assay after 10 min of Cas12a/crRNA sensing time. (B) Sensitivity of 69/70 deletion identification using the CORDS fluorescence reporting assay after 15 min of Cas12a/crRNA sensing time. (C) Schematic representation of Cas12a lateral flow strips to differentiate mutation-containing SARS-CoV-2 variants. (D, E) Sensitivity of 69/70 deletion and N501Y mutation detection using the CORDS fluorescence assay. W: crRNA-W; M: crRNA-M; BC: no target.

After validating the Cas12a fluorescence assay to differentiate mutations, we introduced PCR amplification to improve the sensitivity of mutation identification. The limit of detection (LOD) can reach $10^{-16}$M when Y501-MT or N501-WT amplicons serve as substrates (Figs. 2A, S2, S4A, and S4B). The 69/70 deletion can be identified even when substrate is present at $10^{-17}$ M using TTC as the PAM (Figs. 2B, and S3). Thus, we developed CORDS fluorescence reporting system that rapidly detects mutations from SARS-CoV-2 variants, with high sensitivity and specificity.

The system was made more convenient with reduced dependence of instruments when we combined CORDS with lateral flow strips as a substitute for the fluorescence intensity readout (Fig. 2C). In this assay, 5′-digoxin and 3′-biotin-labeled 14-nt ssDNA and lateral flow strips were used to report the collateral activity of Cas12a (Fig. 2C). Similar to the fluorescence reporting assay, we detected serial concentrations of substrate to determine the sensitivity of the CORDS lateral flow strips reporting system. Results indicated that LOD can reached $10^{-15}$M with Y501-MT or N501-WT as the substrate (Fig. 2D and S4C). This sensitivity can also be achieved for 69/70-MT as the substrate (Fig. 2E). CORDS paper strips system can rapidly detect SARS-CoV-2 variants mutations, with high sensitivity and without the need for a signal detection instrument.

The established CORDS fluorescence and paper strip systems can be used for rapid and sensitive detection of SARS-CoV-2 mutations with DNA targets reversely transcribed from SARS-CoV-2 RNA.

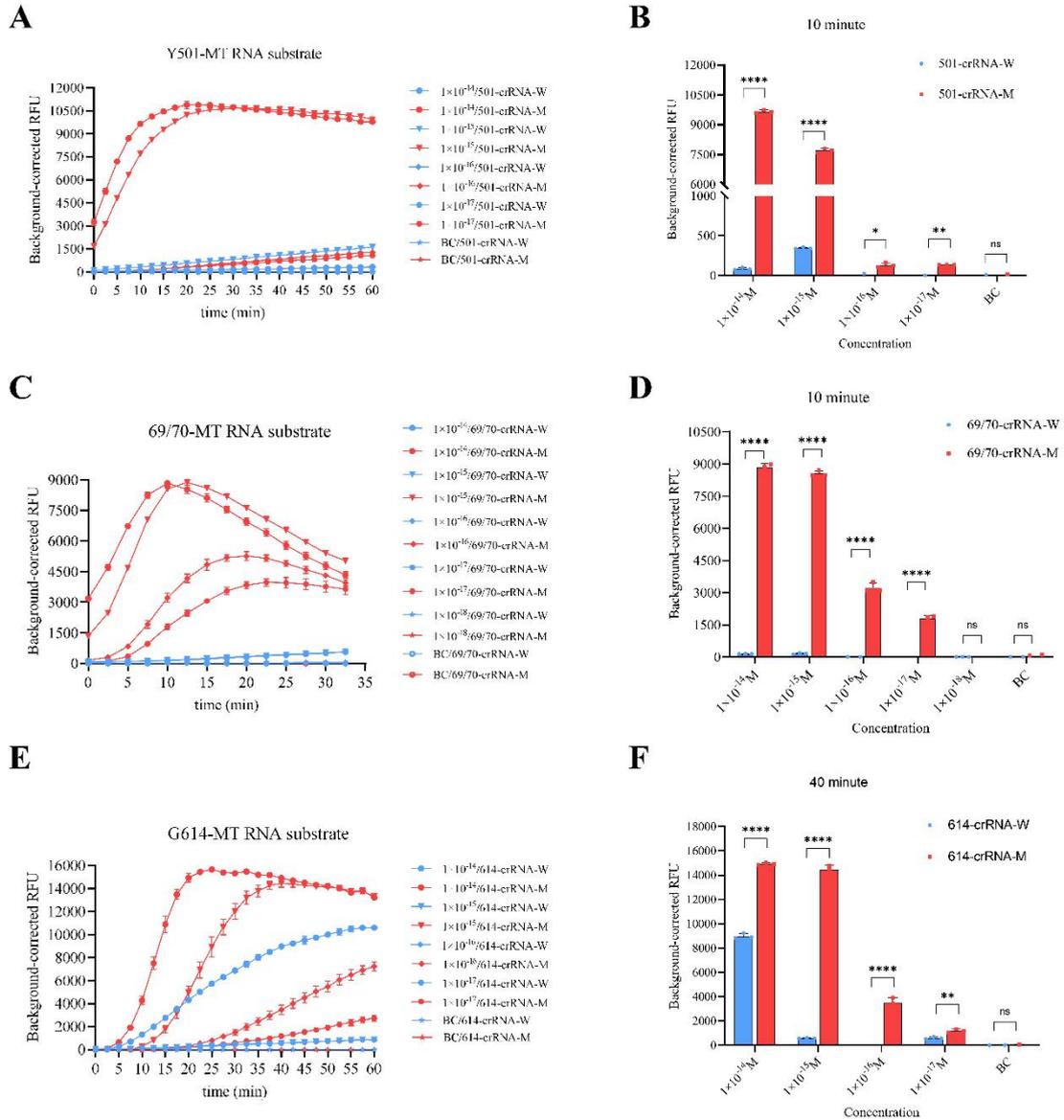

Fig. 3. Sensitivity of SARS-CoV-2 mutant RNA identification using the RT-CORDS fluorescence reporting system. (A, B) Sensitivity of synthetic Y501-MT RNA detection using the RT-CORDS fluorescence assay. (A) Fluorescence kinetics of FQ-ssDNA reporter transcleaved by Cas12a using various Y501-MT RNA concentrations. (B) Background-corrected RFU of Y501-MT detection using various Y501-MT RNA concentrations at 10 min. (C, D) Sensitivity of synthetic 69/70-MT RNA detection using the RT-CORDS fluorescence assay. (E, F) Sensitivity of synthetic G614-MT RNA detection using the RT-CORDS fluorescence assay.

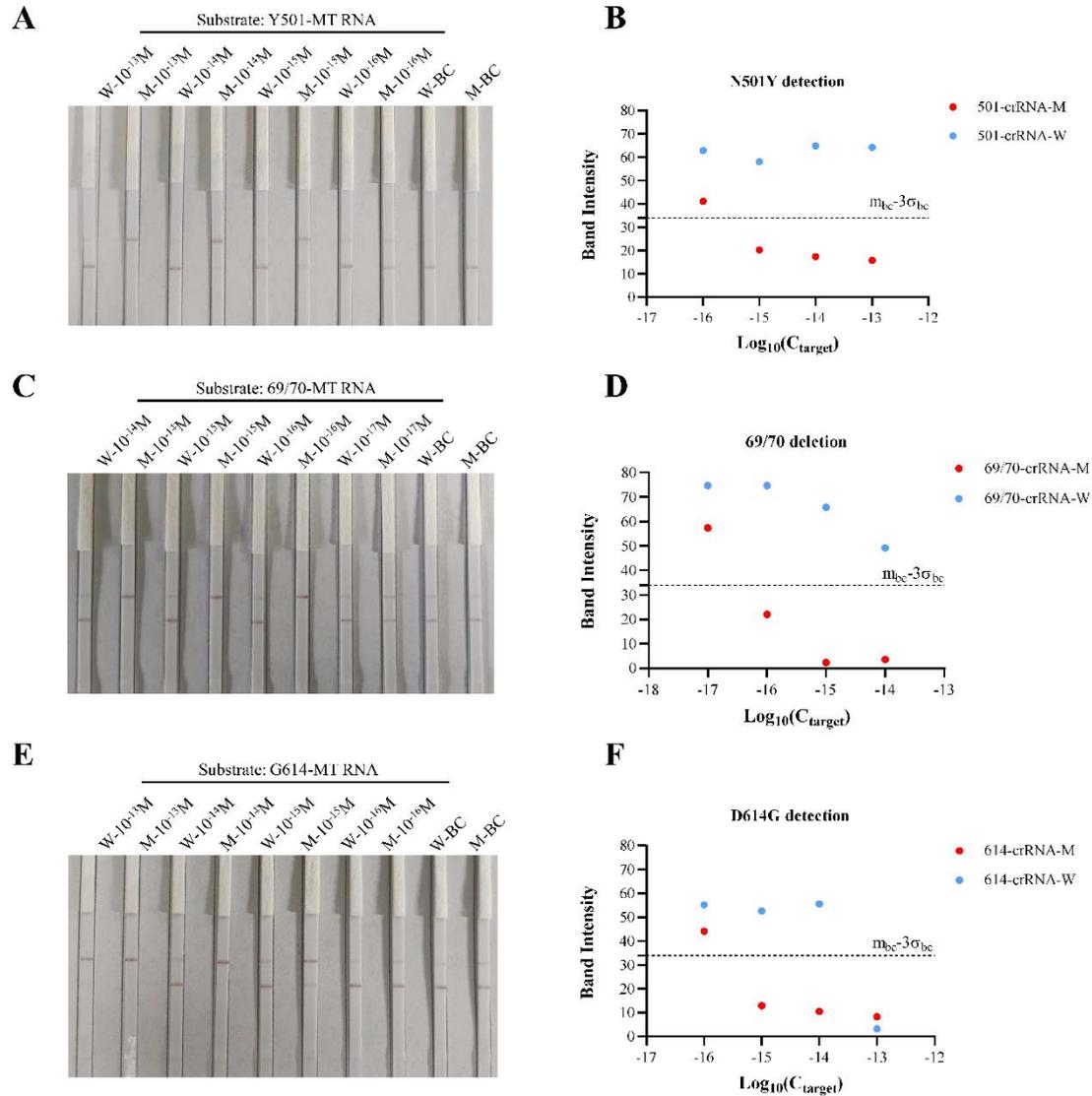

Fig. 4. Sensitivity of SARS-CoV-2 mutant RNA identification using the RT-CORDS lateral flow strip reporting system. (A, B) Sensitivity of N501Y detection using the RT-CORDS lateral flow strips system. (C, D) Sensitivity of 69/70 deletion detection using the RT-CORDS lateral flow strips system. (E, F) Sensitivity of D614G detection using the RT-CORDS lateral flow strips system. (A, C, E) Visual readout of the paper strips. (B, D, F) Mean gray values of the test band at different target concentrations. The dashed line mbc-3σbc indicates the positive cutoff.

## 3.3 RT-CORDS can sensitively and accurately detect the synthetic RNA targets of SARS-CoV-2 mutations

In RT-CORDS, RT-PCR was introduced for RNA reverse transcription and amplification instead of isothermal amplification because the latter is always unstable when combined with reverse transcription (Fig.S5). LOD reached $10^{-17}$ M with Y501-MT RNA as substrate (Figs. 3A, 3B). The sensitivity of the RT-CORDS fluorescence reporting system was higher than that of the CORDS system, perhaps because of the different amplification ability of the DNA polymerase. The sensitivity of RT-CORDS for N501Y detection was $10^{-15}$ M, the same sensitivity as that of CORDS, when lateral flow strips were used as final read out (Figs. 2D, 4A, and 4B). RT-CORDS fluorescence assay for 69/70 deletion identification showed a sensitivity of $10^{-17}$ M, the same value as that achieved for N501Y (Figs. 3C and 3D). The LOD for RT-CORDS lateral flow strip was $10^{-16}$ M (Fig. 4C and 4D), higher than observed for N501Y detection and in a previous study (Bai et al. 2019). The sensitivity for D614G detection was $10^{-17}$ M for RT-CORDS fluorescence and $10^{-15}$ M for RT-CORDS paper strip reporting systems (Figs. 3E, 3F, 4E, and 4F). Increased fluorescence and test band disappearance were observed for 614-crRNA-W at $10^{-14}$ M and $10^{-13}$ M for G614-MT RNA in the fluorescence reporting and lateral flow strip systems. D614G detection is thus more suitable for samples when SARS-CoV-2 viral load is <$10^{-14}$ M (approximately 6000 copies/µL), and appropriate dilution of samples should be considered for higher viral loads. Notably, kinetic analyses showed N501Y and 69/70 deletion can be identified in as small duration as 5–10 min, indicating a shorter reaction time for mutation tracking (Figs. S6 and S7). However, 40 min is the most advisable for D614G detection (Fig. S8).

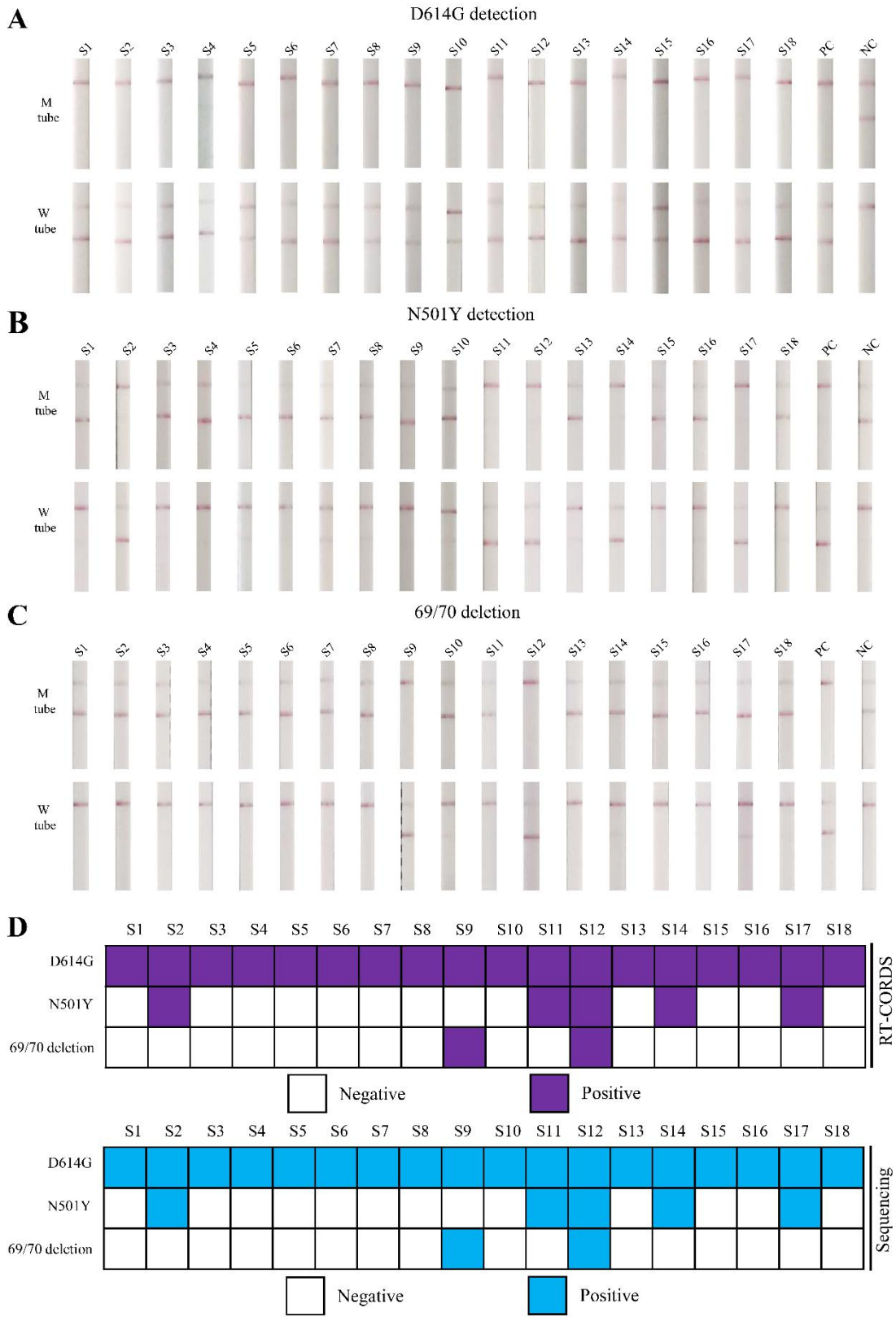

Fig. 5. Detection of SARS-CoV-2 variants using RT-CORDS. (A) D614G detection of SARS-CoV-2 variants using the RT-CORDS paper strip assay. (B) N501Y detection of SARS-CoV-2 variants samples using the RT-CORDS paper strip assay. (C) 69/70 deletion detection of SARS-CoV-2 variants samples using the RT-CORDS paper strip assay. (D) Summary of the detection of three SARS-CoV-2 variant mutations using RT-CORDS and sequencing.

**3.4 RT-CORDS ensures 100% accurate detection of SARS-CoV-2 variants.**

In total, 18 SARS-CoV-2 variant RNA samples were tested using the RT-CORDS paper strip system. All samples carried the D614G mutation (Fig. 5A). Furthermore, five samples carrying the N501Y mutation and two samples carrying the 69/70 deletion were identified (Figs. 5B–5D). Results were 100% consistent with the sequencing data (Figs. 5, and S9–S11).

We simplified the RT-CORDS procedure for storage and use through the lyophilized paper strip system (Fig. S12A). Lyophilization did not alter the LOD for mutation identification (Fig. S12B). RT-CORDS is a rapid, robust, and accurate method for the identification of SARS-CoV-2 N501Y, D614G, and 69/70 deletion mutations and may be used for additional tracking of spike mutations through a simple process in clinical diagnostics (Figs. 6 and S13).

**4. Discussion**

DNA sequencing is the gold-standard for SARS-CoV-2 variant tracking as it can provide complete information regarding the genome sequence. However, sequencing is time-consuming, expensive and inconvenient for general laboratories. Therefore, we established the RT-CORDS for SARS-CoV-2 variant tracking, which is cost-effective, rapid, specific, and sensitive. The system does not require any special laboratory equipment.

High accuracy is the key critical benefit of the RT-CORDS. By introducing a skillful point mutation at the seed region of crRNA, the RT-CORDS can detect the SARS-CoV-2 single-base mutations (N501Y and D614G) with high specificity. RT-CORDS can also accurately identify a small deletion mutation in SARS-CoV-2 (69/70 deletion) with specific crRNAs without introducing a point mutation. In this study, 18 SARS-CoV-2 variants were detected, and all three mutations identified by RT-CORDS were perfectly consistent with the sequencing data, indicating 100% accuracy (Fig. 5). Compared with similar CRISPR-based methods, the accuracy in this study was higher than that of a Cas9-based N501Y detection method (accuracy: 86%) (Kumar et al. 2021), and was consistent with the accuracy of a symRNA-Cas12a D614G detection method (Meng et al. 2021b). Conversely, the accuracy of the most frequently used RT-qPCR for SARS-CoV-2 Marseille-4 variant detection, a point mutation in ORF1, with a specific probe is 93% (Bedotto et al. 2021). Furthermore, the ability of RT-qPCR to differentiate mutations distinguishment is influenced by adjacent mutations, resulting in decreased accuracy (Boudet et al. 2021; Zelyas et al. 2021). This issue illustrates a limitation of RT-qPCR for variant screening.

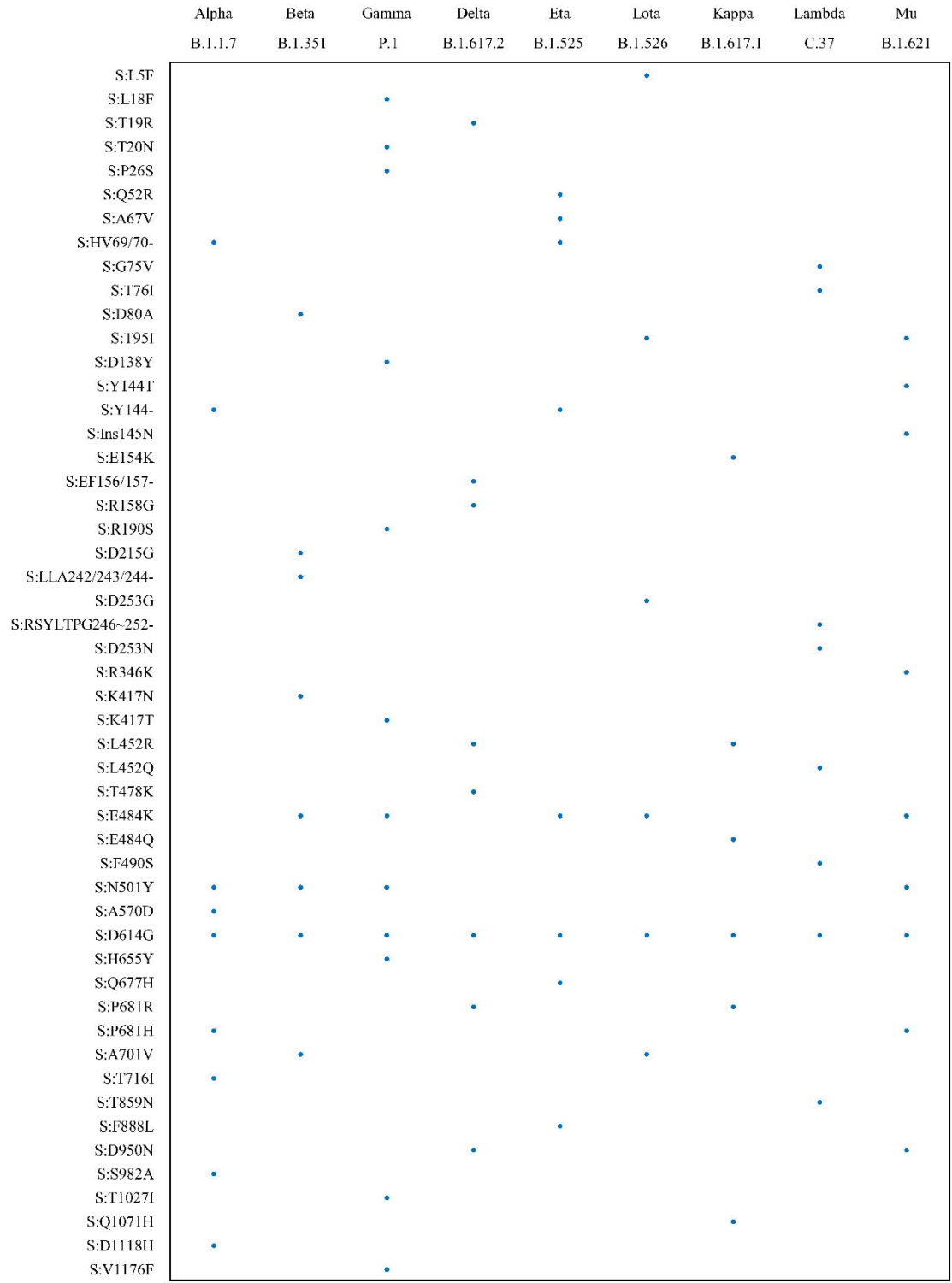

Fig. 6. Detectable mutations which have been reported in spike protein of VOC and VOI by RT-CORDS. Sequences around mutations containing TTTV or TTV PAM sequence were detectable by this method.

The RT-CORDS differentiates mutations in SARS-COV-2 variants accurately and exhibit high sensitivity for all three SARS-COV-2 mutation targets. RT-CORDS with fluorescence showed LODs for all three mutations reached $10^{-17}$ M (approximately 6 copies/µL) (Fig. 3), which was higher than that of the Cas13-based N501Y detection system (100 copies/µL) (de Puig et al. 2021) and was consistent with that reported for a similar Cas12a-based D614G detection method (Huang et al. 2021; Meng et al. 2021b). The RT-CORDS with a visual readout system showed an LOD of $10^{-15}$ M for N501Y and D614G detection and $10^{-16}$ M for 69/70 detection (Fig. 4). The LOD of RT-CORDS for SARS-CoV-2 variant tracking was consistent with our previous study, which detected ASFV using CORDS (Bai et al. 2019). Thus, CORDS detection is robust and reproducible. Clinical data reveal that viral loads of SARS-CoV-2 in throat swab and sputum samples peak around 5–6 days after symptom onset and range from $10^4$ to $10^7$ copies per mL; even higher numbers have been reported for the variants (Fajnzylber et al. 2020; Luo et al. 2021; Pan et al. 2020). RT-CORDS is, therefore, sufficiently sensitive to facilitate the early screening of SARS-CoV-2 variants.

The narrow range of targets makes the PAM sequences a key obstacle for designing an efficient crRNA for mutation detection. We used TTC as an alternative PAM for 69/70 deletion. No TTTV sequence was present near the 69/70 site. Although several studies have reported a lower affinity of TTV for Cas12a loading, the LOD for the 69/70 deletion was the same as that for canonical PAM, indicating that TTV can also be efficient for activating Cas12a system sensitively (Fig. 2B, 3C, and 4C) (Yamano et al. 2017). Sequences adjacent to the *S* gene mutations of SARS-CoV-2 VOC and VOI revealed TTTV and TTV near all mutations, indicating that these mutations are targetable and can be tracked using RT-CORDS (Fig. 6). Conversely, RAY, a Cas9-based SARS-CoV-2 variant tracking system, only targets a subset of mutations reported in SARS-CoV-2 variants, because no NGG PAM sequence exists near the mutation targets (Kumar et al. 2021). Hence, RT-CORDS is widely applicable for mutation tracking.

Despite the single-base resolution of Cas12, the kinetics of CRISPR enzyme for molecular diagnosis recently revealed that realistic LODs for Cas13/Cas12-based detection system were only 10 pM to 100 fM, regardless of the number of targets (Ramachandran and Santiago 2021). Thus, signal amplification is essential either upstream or downstream of CRISPR reactions to achieve high sensitivity (Chen et al. 2020; Li et al. 2021a; Qiao et al. 2021). Isothermal amplification such as RT-RPA or RT-LAMP is a common choice for Cas12 based point-of-care testing (Broughton et al. 2020; Meng et al. 2021b). However, isothermal amplification usually exhibits low efficiency and stability in mutation detection as optimal primer design is limited within the narrow range of mutant targets. Thus, we turned to conventional and stable RT-PCR before Cas12 sensing. Conveniently, RT-PCR is also much cheaper and more available, which facilitate its use in conventional laboratory conditions.

5. Conclusions

In summary, we report that the RT-CORDS is an accurate and sensitive system for identifying single-base mutations (N501Y and D614G) and a small deletion (69/70 deletion) mutation in SARS-CoV-2 variants using CRISPR/Cas12a. This system is easy to assemble for SARS-COV-2 mutation screening in conventional laboratory conditions. This characteristic will be helpful for monitoring the VOC and allow the rapid tracing of epidemic trends. Although RT-CORDS still relies on target amplification as many POCT methods, the high specificity and accuracy of this system make it an ideal choice for high throughput SARS-CoV-2 variant screening.

**Credit authorship contribution statement**

Lizhen Huang and Hongli Du conceived and designed the study, supervised the study, wrote manuscript and acquired the research funds. Changsheng He, Cailing Lin, and Guosheng Mo designed and performed experiments and wrote the manuscript. Binbin Xi and Dongchao Huang performed bioinformatic analysis of sequences. Yanbin Wan, Feng Chen, and Qinxia Zuo analyzed the data. Wanqing Xu and Dongyan Feng made figures. An'an Li, Yufeng Liang, and Guanting Zhang performed variant RNA samples assay. Changwen Ke advised on experimental design.

**Declaration of competing interest**

The authors declare the following financial interests/personal relationships which may be considered as potential competing interests: three patents have been filed through South China University of Technology related to this work.

**Acknowledgments**

This work was supported by the National Natural Science Foundation of China [grant number 31871292]; Fundamental Research Funds for the Central Universities [grant number 2019MS089] from South China University of Technology; Natural Science Foundation of Guangdong Province of China [grant number 2019A1515010619]; Medical Scientific Research Foundation of Guangdong Province of China [grant number A2019432]; The National Key R&D Program of China [grant number 2018YFC0910201].

Supplementary information for

# Rapid and Accurate Detection of SARS-CoV-2 Mutations using a Cas12a-based Sensing Platform

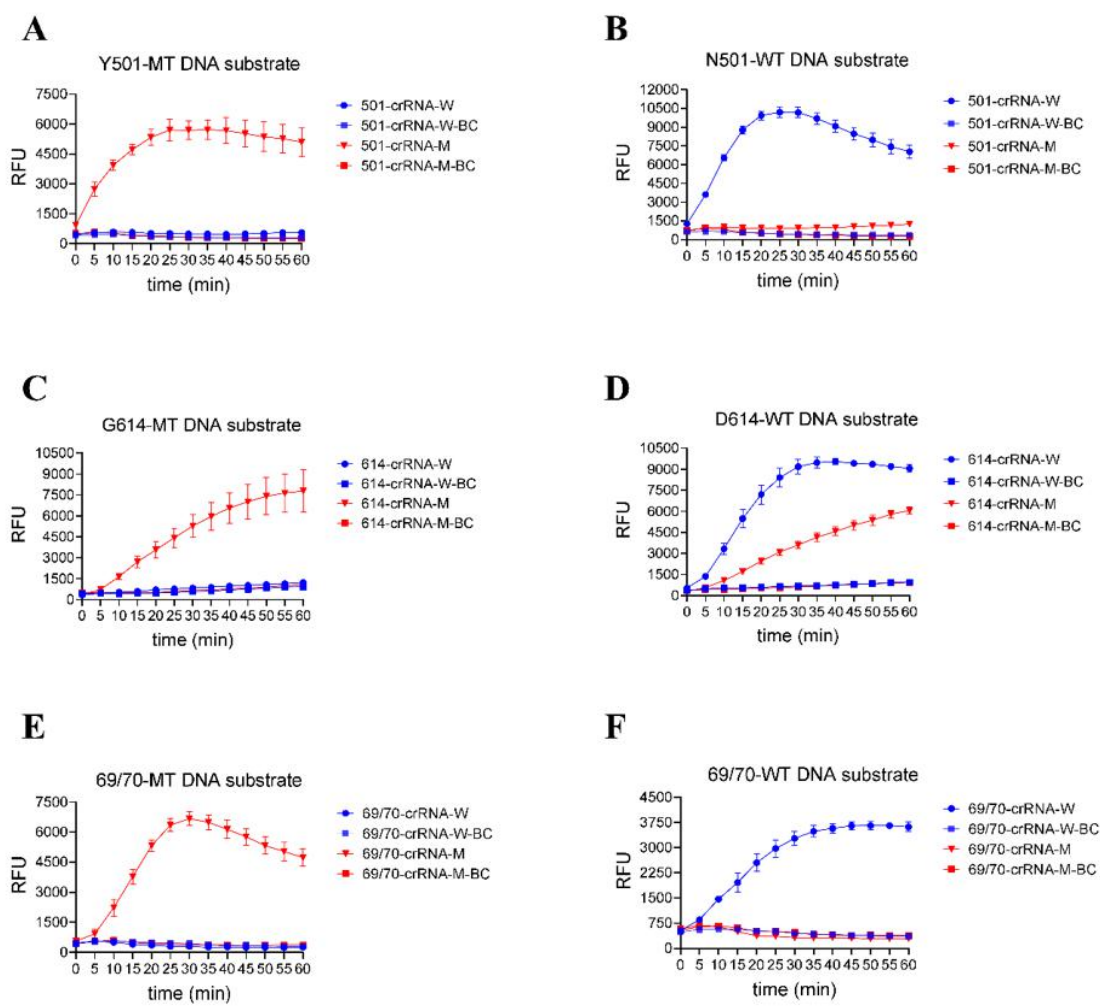

Fig. S1. Identification of N501Y, D614G and 69/70 deletion target through CRISPR/Cas12a trans-cleavage of fluorescence reporter with various reaction time. (A, B) Y501 and N501 identification by Cas12a/crRNA-W and crRNA-M. (C, D) G614 and D614 identification by Cas12a/crRNA-W and crRNA-M. (E, F) 69/70-MT and 69/70-WT identification by Cas12a/crRNA-W and crRNA-M. Link to Fig. 1E to 1G. MT: mutant; WT: wild-type. BC: no target.

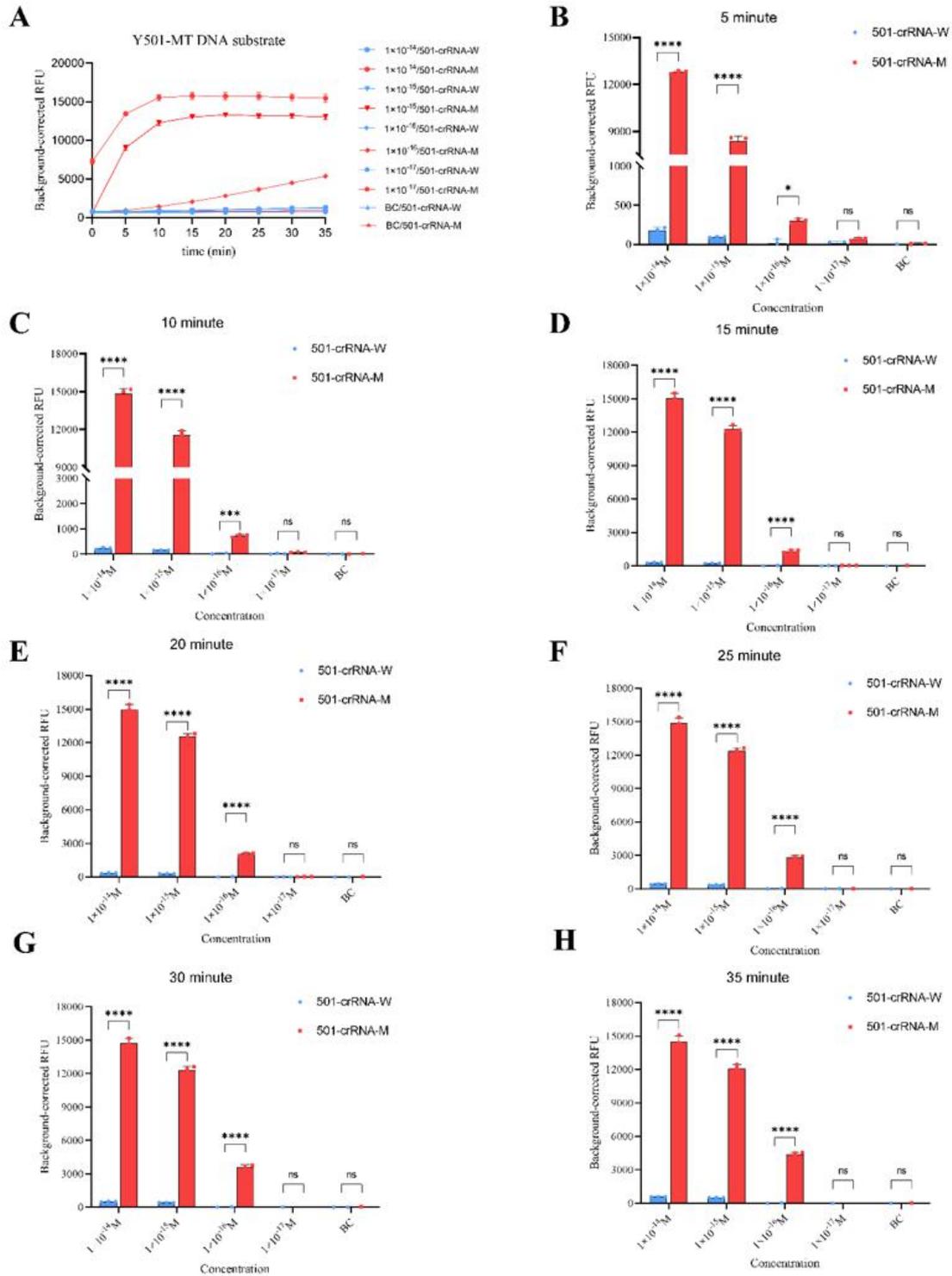

Fig. S2. (A) Fluorescence kinetics of FQ-ssDNA reporter trans-cleaved by Cas12a with various Y501-MT DNA concentrations. (B to H) Sensitivity of Y501-MT detection using CORD fluorescence reporting system at different

reaction time.

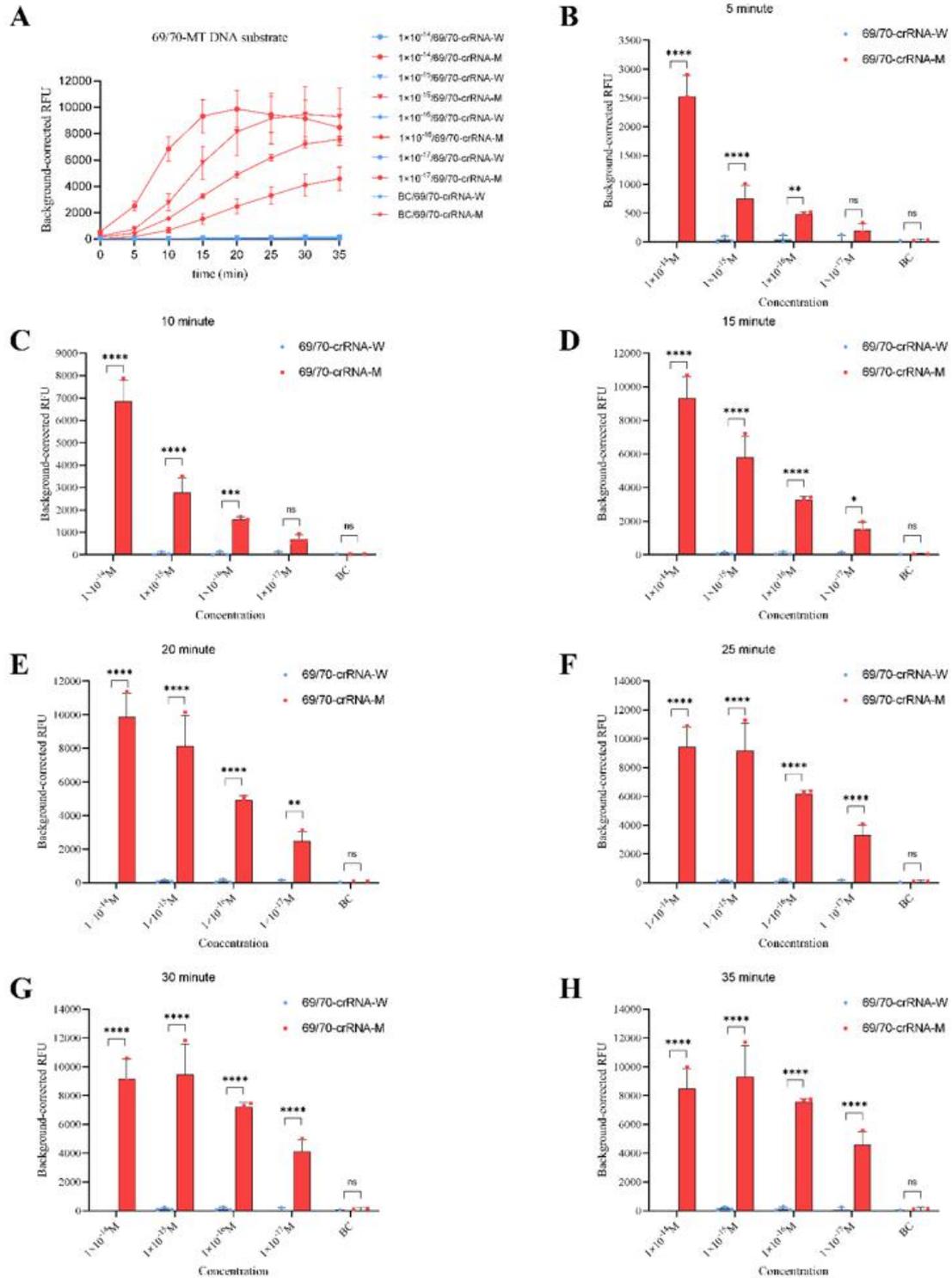

Fig. S3. (A) Fluorescence kinetics of FQ-ssDNA reporter trans-cleaved by Cas12a with various Y501-MT DNA

concentrations. (B to H) Sensitivity of Y501-MT detection using CORD fluorescence reporting system at different reaction time.

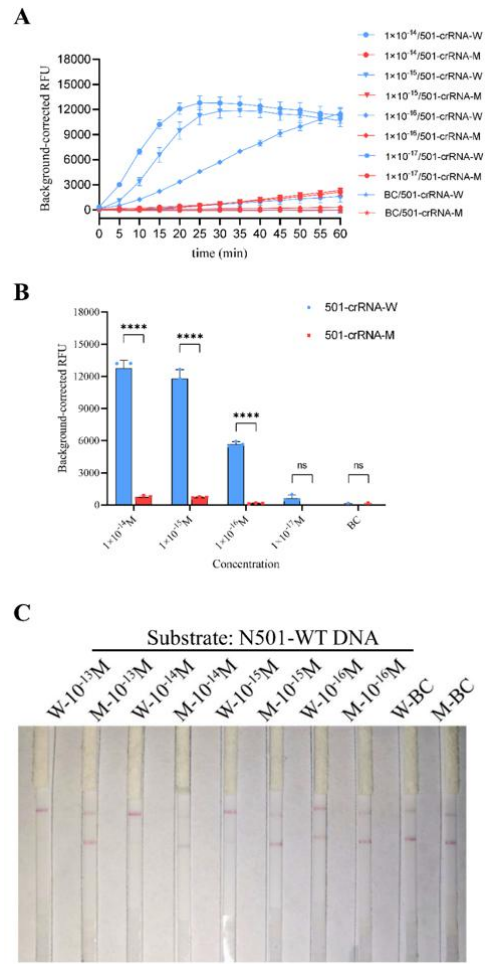

Fig. S4. Sensitivity of N501-WT detection using CORDS. (A) Fluorescence kinetics of FQ-ssDNA reporter trans-cleaved by Cas12a with various N501-WT DNA concentrations. (B) Background-corrected RFU of N501-WT detection with various N501-WT DNA concentrations at 30 min. (C) Sensitivity of N501-WT detection by CORD paper strips system. W: 501-crRNA-W; M: 501-crRNA-M.

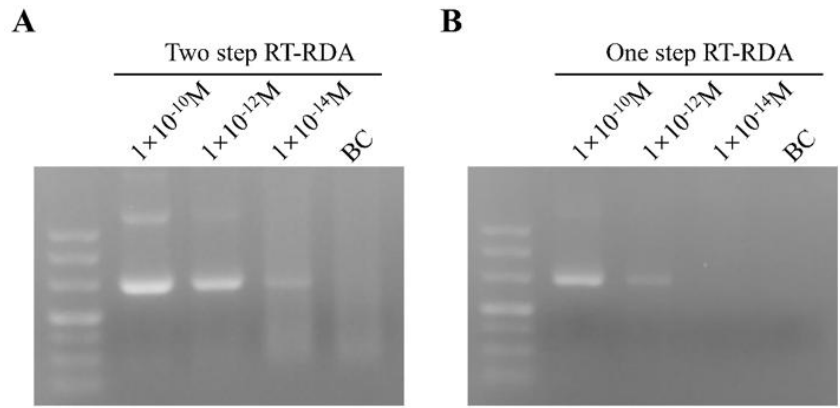

Fig. S5. Isothermal amplification of D614G RNA target by RT-RDA (Magigen Biotech, China). (A) two step RT-RDA amplification, reverse transcription and amplification were performed individually. (B) one step RT-RDA amplification, reverse transcription and amplification were performed in one-pot simultaneously.

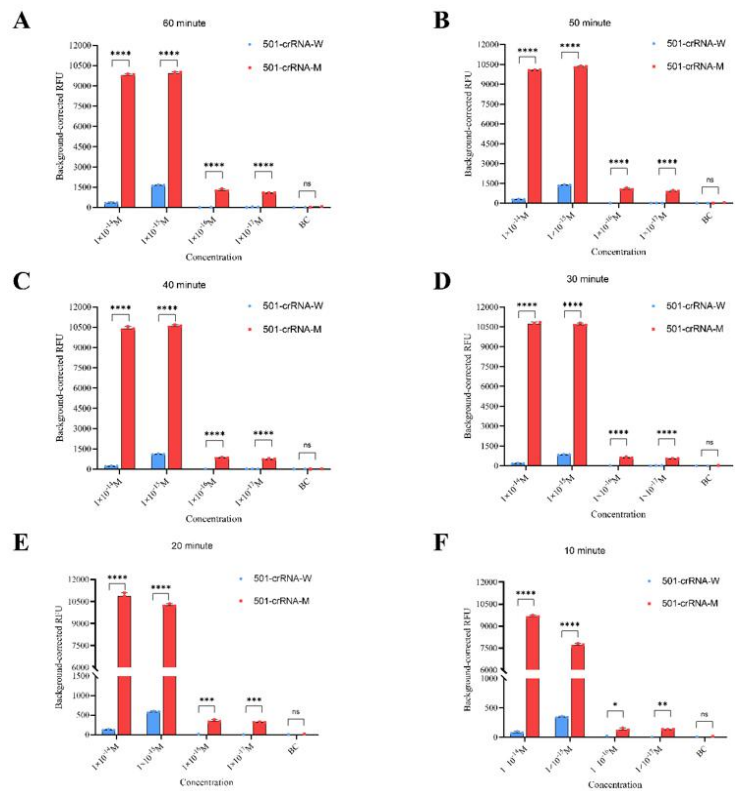

Fig. S6. Sensitivity of N501Y detection by RT-CORDS fluorescence reporting assay at different reaction time.

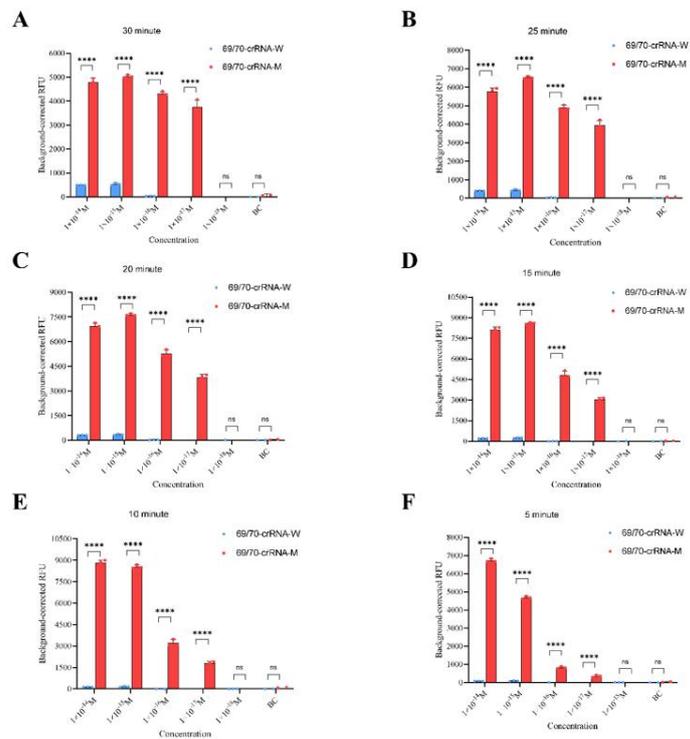

Fig. S7. Sensitivity of 69/70 deletion identification by RT-CORDS fluorescence reporting assay at different reaction time.

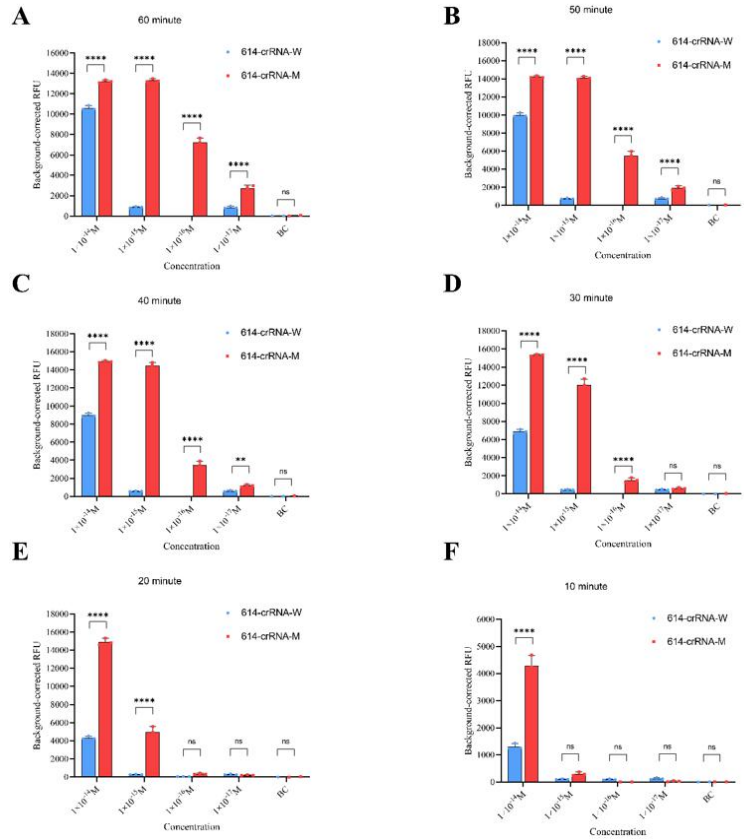

Fig. S8. Sensitivity of D614G detection by RT-CORDS fluorescence reporting assay at different reaction time.

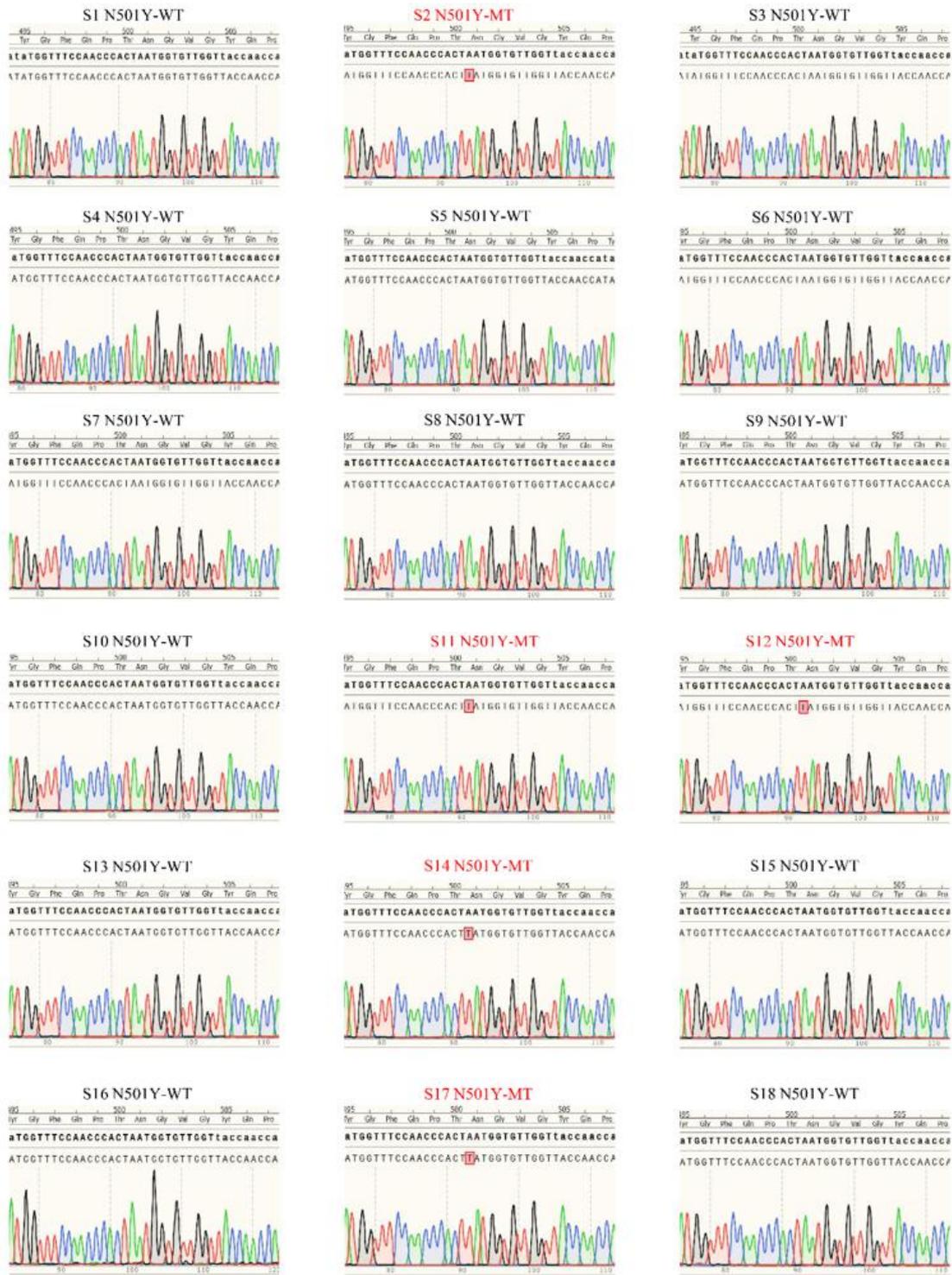

Fig. S9. Further N501Y validation of SARS-CoV-2 variants samples by sequencing.

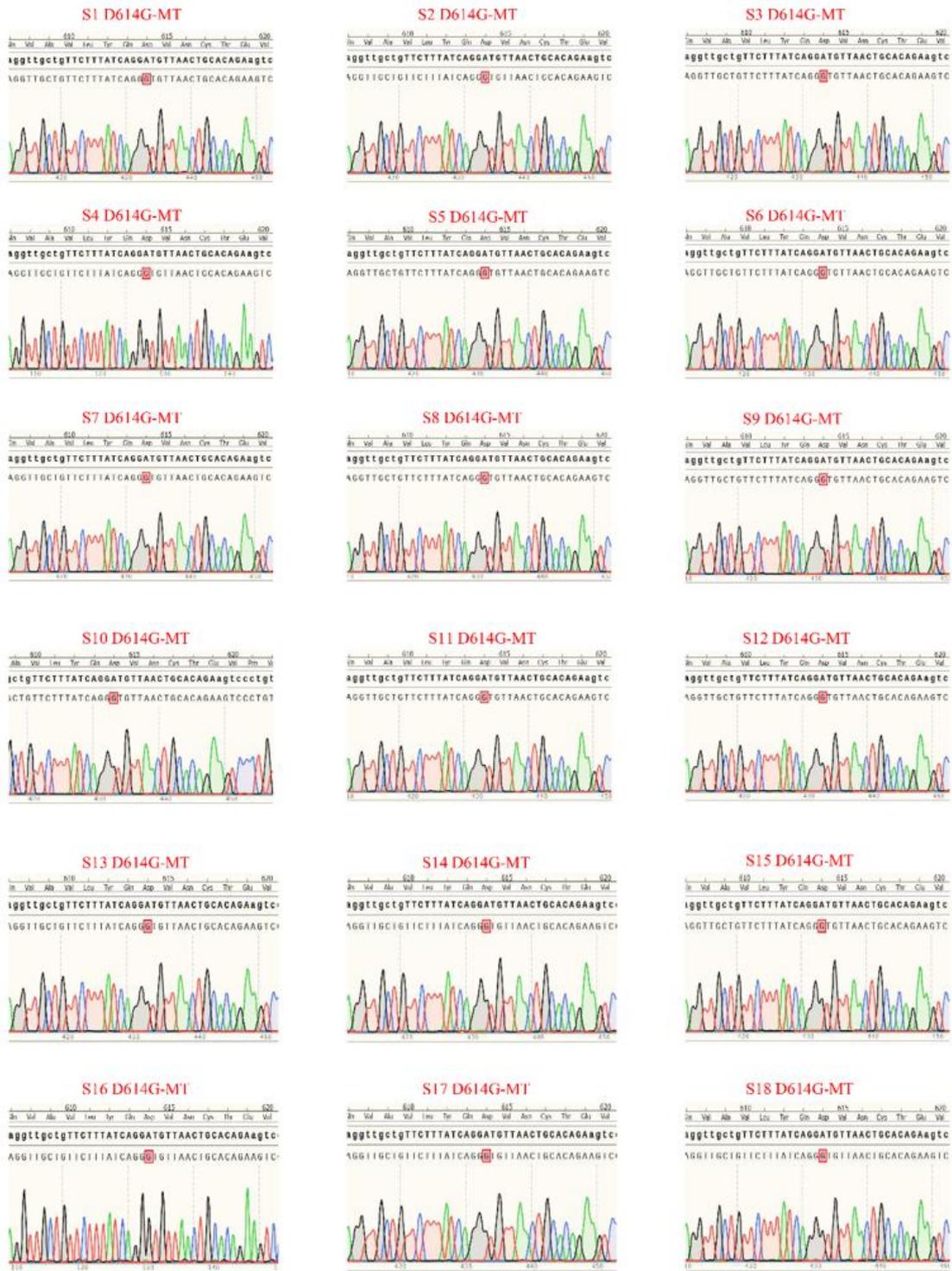

Fig. S10. Further D614G validation of SARS-CoV-2 variants samples by sequencing.

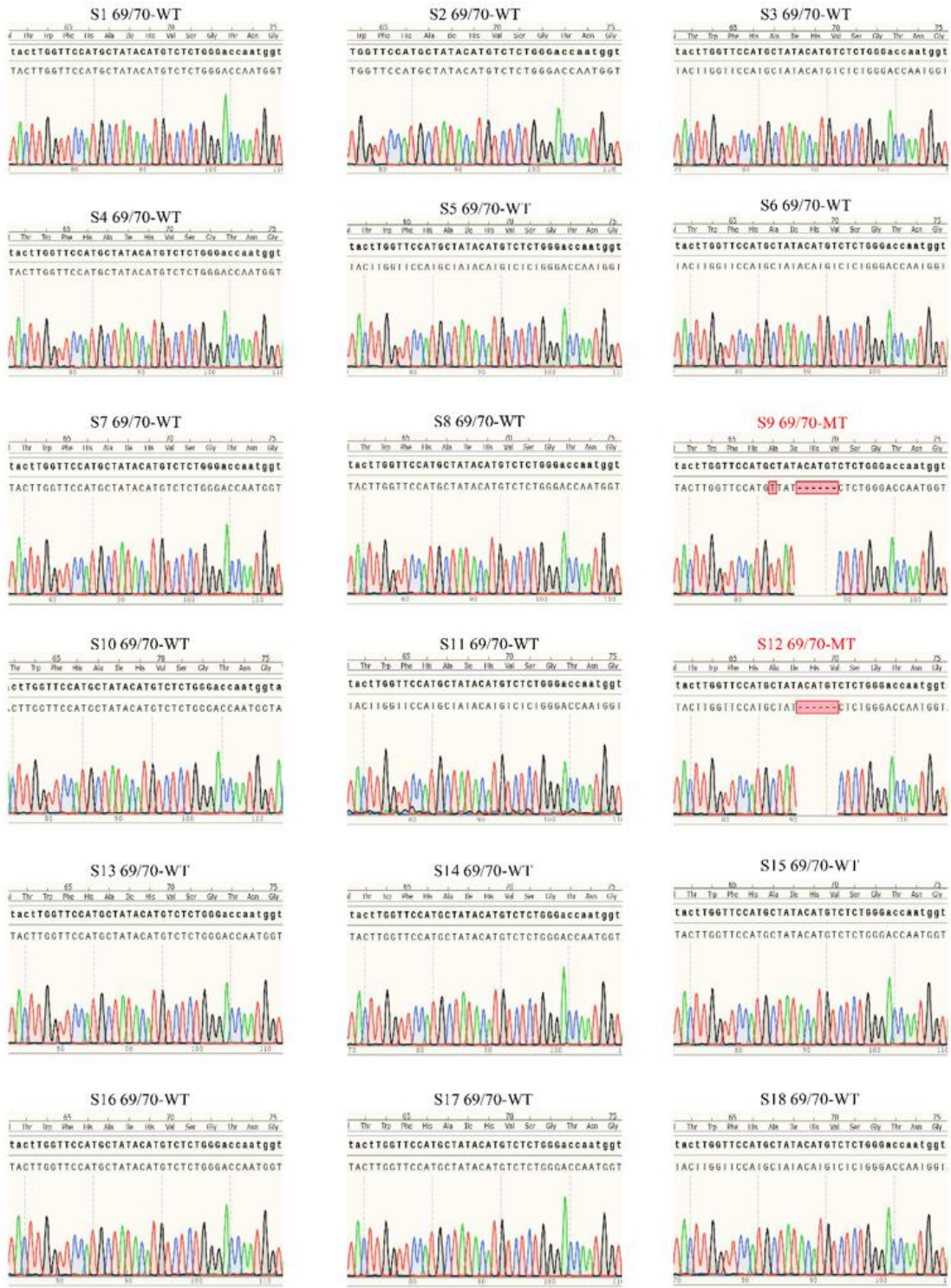

Fig. S11. Further 69/70 deletion validation of SARS-CoV-2 variants samples by sequencing.

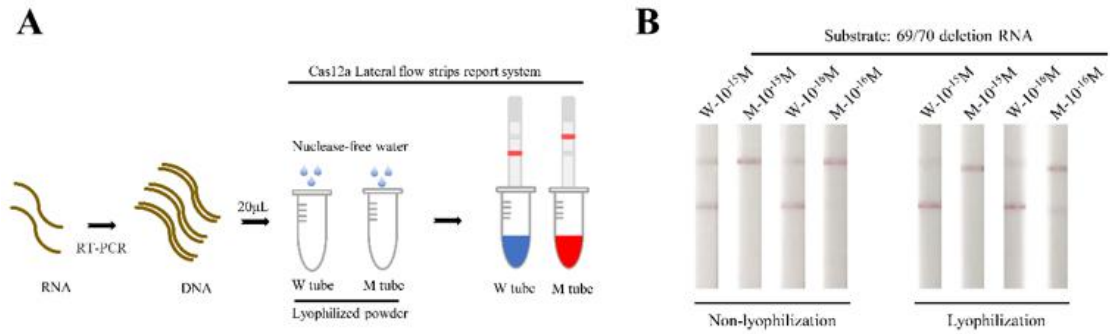

Fig. S12. Lyophilization of RT-CORDS lateral flow strips reporting system. (A) process of mutation identification by lyophilized RT-CORDS paper strips assay. (B) Sensitivity of lyophilized RT-CORDS paper strips assay.

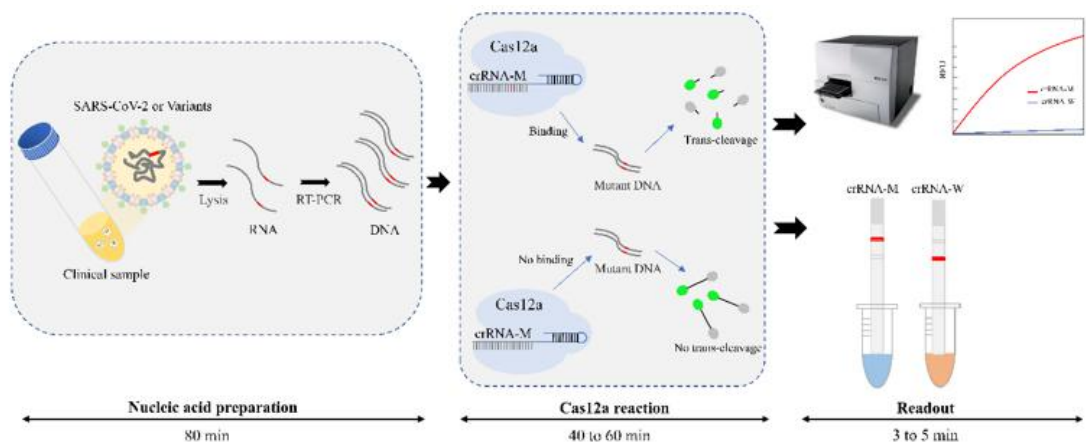

Fig. S13. Process for RT-CORDS to identify SARS-CoV-2 variants.